\documentclass[onecolumn,prb]{revtex4}

\usepackage{graphicx}
\usepackage{dcolumn}
\usepackage{bm}
\begin{document}

\preprint{APS/version 0.1}

\title{Schwinger term of neutron Hamiltonian measurable by polarization change in a spin-echo spectrometer}

\author{Victor Otto de Haan}
 \affiliation{BonPhysics B.V., Laan van Heemstede 38, 3297 AJ, Puttershoek, The Netherlands}
\date{\today}

\begin{abstract}
The Schwinger term of the neutron Hamiltonian is due to the effect of an electric field on the moving neutron. Although this effect is extremely small, it can be measured by the novel spin-echo interferometers. 
\end{abstract}

\maketitle 
\section*{Introduction}
The current limit on the charge of the neutron is about $10^{-40}$~C. The limit on its electric dipole moment is equally very small $10^{-46}$~Cm~\cite{Nico2005}. Hence, one could expect that the neutron is for all practical purposes a neutral particle without any electric dipole moment. Hence, for a stationary neutron the interaction with an electric field is non-existent or in any case immeasurably small. 
However, the neutron is not stationary. It is a particle moving with a finite velocity. 

Let a neutron move with particle velocity, $v$ with respect to a laboratory frame of reference. A static electric field stationary in the laboratory frame results in a magnetic field in the reference frame of the neutron. Vice versa, a static magnetic field stationary in the laboratory frame results in an electric field in the reference frame of the neutron. However, the electric field in the reference frame of the neutron is of no interest as the neutron is neutral and has no detectable electric dipole moment. To calculate the magnetic field in the reference frame of the neutron due to an electric field in the laboratory frame one can use two approaches. One approach is to think of the electric field as built up by the charges on the plate of a virtual capacitor creating the field and use the Minkowski material equations to calculate the magnetic field due to moving charges. The other approach is to use the Poincar\'{e}-Lorentz transformations.

\section*{Moving charged capacitor and Minkowski material equations}
\begin{figure}
\begin{picture}(200,180)
\put(0,70){\scalebox{0.38}{\includegraphics{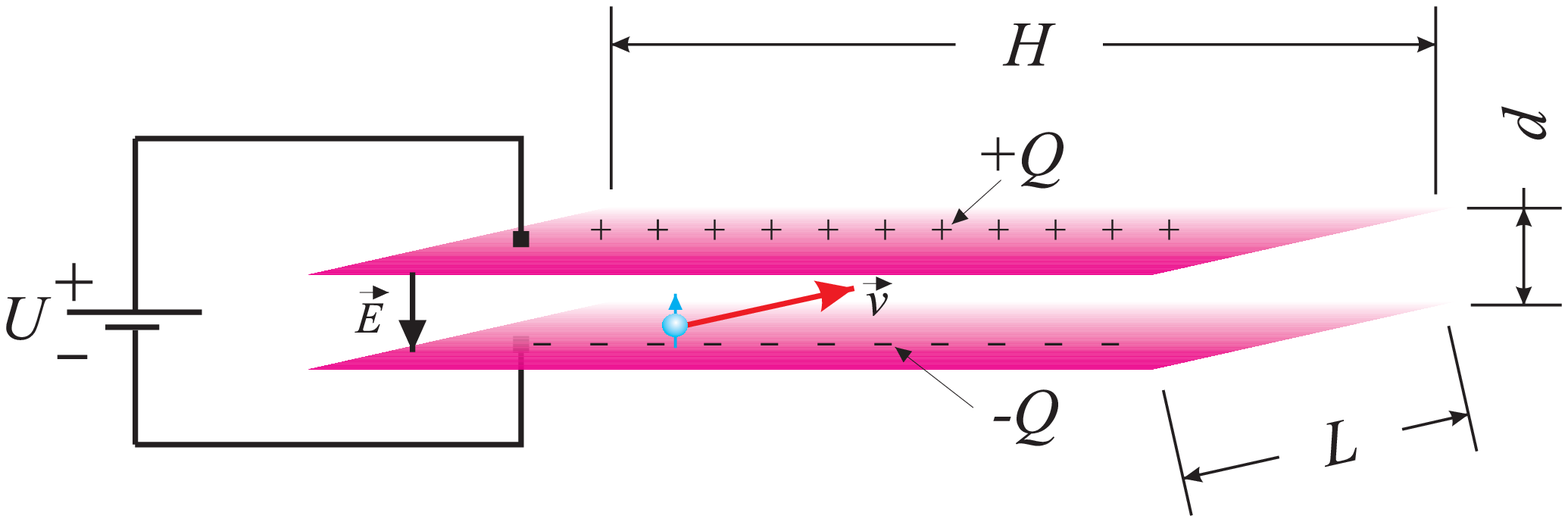}}}
\put(0,153){Laboratory reference frame}
\put(26,0){\scalebox{0.38}{\includegraphics{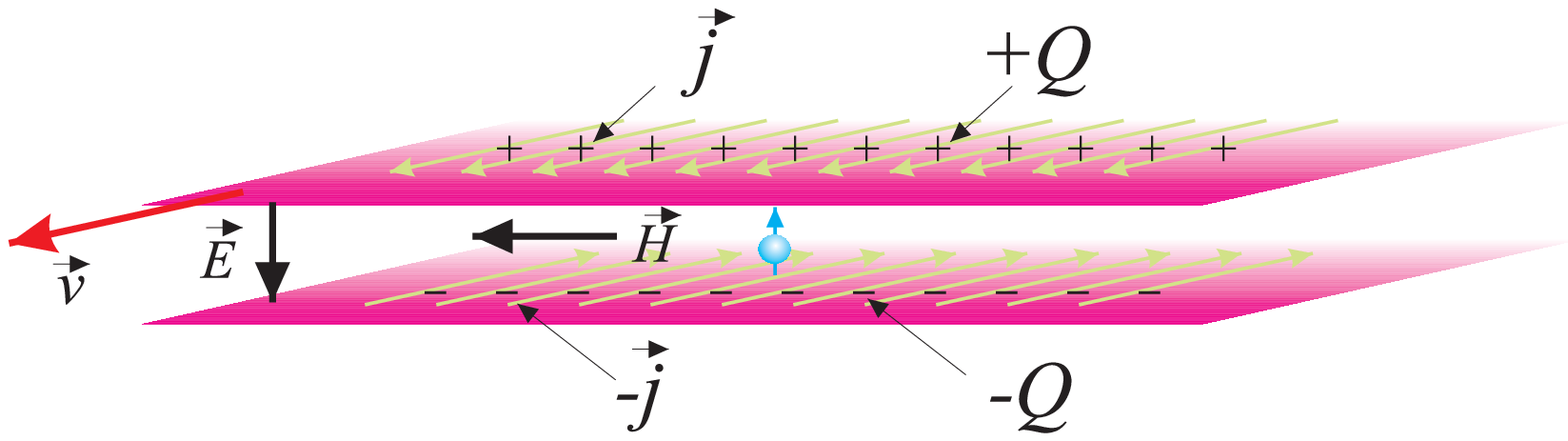}}}
\put(0,53){Neutron reference frame}
\end{picture}
\caption{\label{fig:1} Neutron traveling between the charged plates of a capacitor viewed from laboratory reference frame in which the capacitor is at rest and from a reference frame in which the neutron is at rest, where the charge move along with the plates resulting in a current density on the plates and a magnetic field acting on the neutron.}
\end{figure}

In the laboratory frame of reference a capacitor is charged by an external power supply with a voltage $U$ up to a charge $Q$. This is shown schematically in the top panel of figure~\ref{fig:1}. 
It is assumed that the distance $d$ between the plates of the capacitor is much smaller than its length, $L$ along the neutron path or its width, $H$. The charge, $Q$ is related to the capacitance, $C$ of the capacitor according to $Q=CU$. The capacitance of such a plate capacitor can be calculated from 
\[
C=\frac{\epsilon_r\epsilon_0 O }{d} ,
\] 
where $O=LH$ is the area of the plates, $\epsilon_0$ is the permittivity of free space $= 8.854 187 817... \times 10^{-12}$ C/Vm and $\epsilon_r$ is the relative permittivity of the material between the plates. Hence, the charge on the plates is related to the electric field between the plates as
\[
Q = CU = CEd = \epsilon_r\epsilon_0 O E .
\]
In the reference frame where the neutron is at rest this charge moves with a velocity of $v$ in the direction opposite to the neutrons velocity in the laboratory reference frame, creating a surface current density, $\vec{j}$ depending on the charge density and the velocity:
\[
\vec{j} = \frac{Q}{O}\vec{v} = \epsilon_r\epsilon_0 E \vec{v}.
\]
This surface current density can be regarded as a current, $I$ through an infinite coil resulting in a magnetic field, $\vec{H}$ perpendicular to the electric field and surface current density as indicated in figure~\ref{fig:1}. The strength of the magnetic field can be calculated according to
\[
\oint{\vec{H}(\vec{r}) \cdot \vec{dr}} = I,
\] 
or
\[
H = j.
\] 
Note, that the above equations can be applied regardless the properties of the moving material between the plates. If there is no material in between, the magnetic induction in the reference frame of the neutron due to the static electric field in the laboratory reference frame is
\[
B = \mu_0 H =  \mu_0  \epsilon_0 E v = \frac{E v}{c^2},
\]
where $\mu_0$ is the permeability of free space $= 4\pi \times 10^{-7}$ N/A$^2$.
If a material is present between the plates, the Minkowski material equations for a material moving with velocity $\vec{v}$ must be used to calculate the magnetic induction from the magnetic field. These equations are given by~\cite{MinkowskiMaterialEquations}:
\[
\begin{array}{l}
\vec{D} + \frac{ \vec{v} \times \vec{H}}{c^2} = \epsilon_r \epsilon_0 \left\{\vec{E}+ \vec{v} \times \vec{B}  \right\},
\\
\vec{B} -  \frac{ \vec{v} \times \vec{E}}{c^2}  = \mu_r \mu_0 \left\{\vec{H}-\vec{v} \times \vec{D} \right\},
\end{array}
\]
where $\mu_r$ is the relative permeability of the material between the plates. In the non-relativistic velocity limit these equations reduce to the Galilean electric constitutive equations~\cite{Galileanelectric}
\[
\begin{array}{l}
\vec{D}= \epsilon_r \epsilon_0 \vec{E},
\\
\vec{B} = \mu_r\mu_0\vec{H} - \left(\mu_r\epsilon_r-1\right)\frac{\vec{v}\times\vec{E}}{c^2},
\end{array}
\]
so that again,
\[
B = \frac{ E v}{c^2}.
\]
Here it was used that in the Galilean electric non-relativistic velocity limit the electric field is the same as in the stationary frame.
\section*{Poincar\'{e}-Lorentz transformations}
Under a Poincar\'{e}-Lorentz-Minkowski transformation the magnetic induction transforms as~\cite{PoincareLorentztransformations}
\[
\vec{B'} = \gamma\vec{B}+(1-\gamma)\frac{\vec{v}(\vec{v}\cdot\vec{B})}{v^2}-\frac{\vec{v}\times\vec{E}}{c^2},
\]
where $\gamma=\sqrt{1-v^2/c^2}$. Yielding, in the non-relativistic limit, directly the result obtained in the previous approach.
\section*{Schwinger term of Hamiltonian}
The above result can also be derived directly from the non-relativistic limit of the Dirac equation. The part of the Hamiltonian associated with moving magnetic moment in the electric field is given by the Schwinger term~\cite{Sears1986}
\[
 - \mu_n \widehat{\sigma} \cdot \frac{\vec{E} \times \vec{v}}{c^2},
\]
where $\widehat{\sigma}$ the Pauli spin matrix vector and $\mu_n=-60.308~neV/T$ the neutron magnetic moment.
\section*{Foldy and polarisation terms of Hamiltonian}
There are two other terms in the Hamiltonian expressing interaction with the electric field~\cite{Sears1986}. One of them is the Foldy term
\[
 - \frac{\hbar \mu_n }{2mc} \nabla \vec{E} = - \frac{\hbar \mu_n }{2mc} \sigma,
\]
where $\sigma$ is the charge density. The neutron can be expected to have a non vanishing electric polarisability $\alpha$ as a result of which the neutron will acquire an electric dipole
moment $\alpha \vec{E}$ in the electric field. The polarisation term of the Hamiltonian is
and polarisation term
\[
 - \frac{1}{2}\alpha E^2 .
\]
These two terms do not depend on the spin direction and should not show any effect on the rotation of the polarisation vector.
\section*{Polarisation Rotation}
The rotation of the polarisation vector due to a magnetic field over a length $L$ is given by the Larmor precession
\[
\phi = \gamma_LBt_L,
\]
where $\gamma_L = \mu_n / \hbar$ and $t_L=L/v$ is the travel time of the neutron with wavelength, $\lambda$. For the Schwinger effect the rotation is
\[
\phi_{S} = \frac{\gamma_L}{c^2} E L , 
\]
where $\gamma_L /c^2 \approx 2.0\   \mu$rad/kV. For a minimum detectable change of 0.1~rad and a maximum vacuum electric field of 3.10$^7$ V/m, the minimum length is 1.6~m. Further material examples are given in table~\ref{tab:1}. 
\begin{table}
\begin{tabular}{|l|l|l|l|l|l|l|l|l|} \hline
 Material         & $E_b$    & $\epsilon_r$ & $B_{eff}$ for 0.25 nm & $L_{min}$   & $\alpha $ for 0.25 nm  \\
                  & MV/m		 &              & $\mu$T  & m           &  picorad              \\ \hline 
Air               & 3        & 1            & 0.05    & 16          &  1.3                  \\ 
Vacuum            & 30       & 1            & 0.53    & 1.6         &  13                   \\ 
Fused Quartz      & 25       & 3.8          & 0.44    & 2.0         &  11                   \\  
Silicon           & 60       & 8.5          & 1.06    & 0.7         &  27                   \\
AlN               & 120      & 8.5          & 2.11    & 0.4         &  54                   \\ 
Teflon            & 170      & 2            & 3.0     & 0.3         &  72                   \\  
SiC               & 300      & 9.7          & 5.3     & 0.15        &  134                  \\
\hline
\end{tabular}
	\caption{Breakdown electric field, $E_b$, relative permittivity, $\epsilon_r$, effective magnetic field, $B_{eff}$ for 0.25~nm neutrons and minimum length, $L_{min}$ for a polarisation rotation of 0.1~rad. Further, the diffraction angle $\alpha$ for a wavelength of 0.25~nm.}
	\label{tab:1}
\end{table}
It is clear that the effect is surely detectable, but the flight path trough the electric field must be very long or the phase change must be detected very accurately. It has been measured by Kaiser and Cimmino~\cite{Kaiser1988,Kaiser1989,Cimmino1989} by using a Bonse-Hart interferometer. It also can be used to check the validity of the Minkowski material equations for neutrons moving in a material. However then the flightpath through the material must not be more than a few centimeters.
\section*{Effect enhancement}
\begin{figure}
\begin{picture}(300,140)
\put(0,0){\scalebox{0.65}{\includegraphics{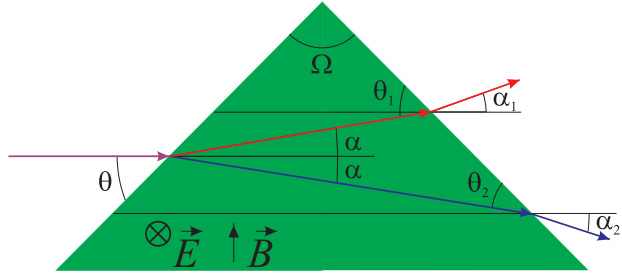}}}
\end{picture}
\caption{\label{fig:2} Neutron splitting due to a special shaped electric field region corresponding to an effective magnetic field region. After exit of the region the neutron wave remains split resulting in a continuous polarisation rotation after transmission through the region.}
\end{figure}

\begin{figure}
\begin{picture}(480,80)
\put(0,0){\scalebox{0.55}{\includegraphics{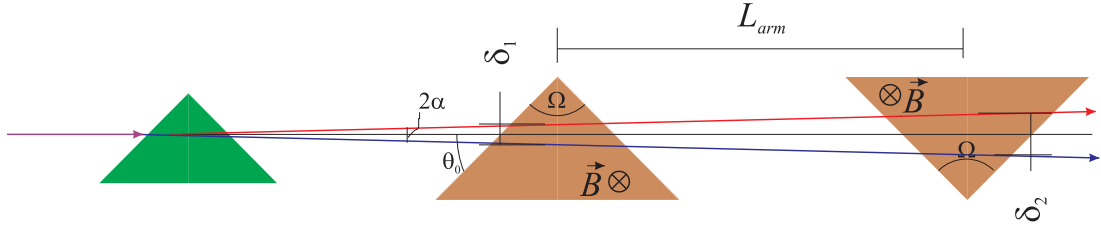}}}
\end{picture}
\caption{\label{fig:2b} Schematic trajectory of spin-up wave function and spin-down wave function through one arm of a spin-echo spectrometer after transmission through the electric field region introducing a splitted wave function.}
\end{figure}

\begin{figure}
\begin{picture}(480,80)
\put(0,0){\scalebox{0.45}{\includegraphics{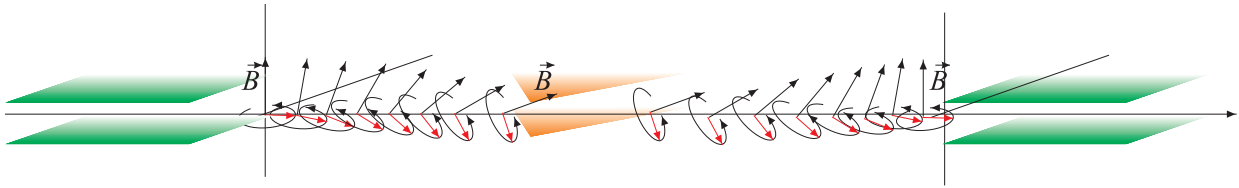}}}
\end{picture}
\caption{\label{fig:3} Schematic view of rotation of plane of polarisation rotation needed in OffSpec.}
\end{figure}

The accuracy of the measurement of the Schwinger term can be greatly enhanced by shaping the electric field region in a similar way as for the magnetic field regions of the spin-echo instruments. This is shown schematically in figure~\ref{fig:2}. Due to the different potential energy the spin-up wave function is diffracted differently from the spin-down function. The refraction angles only depend on the slope of the borders of the field regions~\cite{Keller1997,DeHaanBook}:
\begin{equation}\label{eq:refang}
\alpha = \frac{ m \mu_n B_{eff} }{\hbar^2  k^2} \cot \theta.
\end{equation}
As the refraction angles are extremely small $\theta_1 \approx \theta_2$. In the following it is also assumed that the angle $\Omega$ is chosen in such a way that $\theta_1\approx\theta_2\approx\theta$,  so that $\alpha_1 \approx \alpha_2 \approx \alpha$. In the following the indices on $\alpha$ and $\theta$ are omitted. If the effective magnetic field is substituted by its electric cause, $\alpha$ is given by
\begin{equation}\label{eq:eff}
\alpha= \gamma_L \frac{ \lambda E}{\pi c^2} \cot \theta .
\end{equation}
Let us assume that the wave function incident on the triangle is given by
\[
 \left( \begin{array}{c} \Psi^{+} (\vec{r})  \\ \Psi^{-} (\vec{r}) \end{array} \right) = \frac{e^{ikx}}{\sqrt{2}} \left( \begin{array}{c} e^{i\phi_0/2}  \\ e^{-i\phi_0/2} \end{array} \right)
\]
then inside the triangle the wave function is can be found by matching the wave functions and their derivatives at the transition line $y=x \tan\theta $ given by (ignoring edge effects)
\[
 \left( \begin{array}{c} \Psi^{+} (\vec{r})  \\ \Psi^{-} (\vec{r}) \end{array} \right) = \frac{1}{\sqrt{2}} \left( \begin{array}{c} e^{ik(1+\alpha\tan\theta)(x\cos\alpha - y\sin\alpha)+i\phi_0/2}  \\ e^{ik(1-\alpha\tan\theta)(x\cos\alpha+y\sin\alpha)-i\phi_0/2} \end{array} \right)
\]
and after the triangle by matching the wave functions at the transition line $y=(L-x) \tan\theta $
\[
 \left( \begin{array}{c} \Psi^{+} (\vec{r})  \\ \Psi^{-} (\vec{r}) \end{array} \right) = \frac{e^{ik(x-L)\cos2\alpha}}{\sqrt{2}} \left( \begin{array}{c} e^{ik(1+\alpha\tan\theta)L-iky\sin2\alpha+i\phi_0/2}  \\ e^{ik(1-\alpha\tan\theta)L+iky\sin2\alpha-i\phi_0/2} \end{array} \right),
\]
where refraction angle $\alpha$ is given by
The corresponding polarisation rotation due to the triangle, $\phi_e$ is given by
\[
\phi_e = 2k\alpha (L\tan\theta-2y).
\]
The first term is the Larmor precession in the triangle and the second term is a zero field precession due to the splitting. These equations were derived in a plane wave approximation. Normally the neutron is a wave packet and this should be taken into account. This goes beyond the scope of this derivation, but it can be shown that in such a case the wave function only has an appreciable amplitude within the transversal coherence length of the neutron, which is of the order of several tens of nanometer. Hence, in general the polarisation rotation due to the second term can be ignored.
However, the spin-up wave function and the spin-down wave function are refracted with an opposite angle. Hence, after some distance after the sample a splitting has occurred. When this splitted beam is incident on one arm (with length $L_{arm}$) of a spin-echo spectrometer the spin-up wave and the spin-down wave will traverse the magnetic field regions at different heights. The first region at a height difference of $\delta_1$ and the second one at a height difference of $\delta_2$, where $\delta_2-\delta_1=2\alpha L_{arm}$. This is shown schematically in fig.~\ref{fig:2b}.
Hence, the spin-up wave function effectively traverses a longer distance through the magnetic field regions of the spin-echo arm than the spin-down wave function. The path length difference is given by $2(\delta_2-\delta_1)\cot\theta_0$, where $\theta_0$ is a similar angle as $\theta$ for the spectrometer used. Hence, the extra phase acquired by the spin-up wave function with respect to the spin-down wave function is
\[
\phi_e = \frac{2m\mu_nB}{k \hbar^2}(\delta_2-\delta_1)\cot\theta_0 = \frac{2m\mu_nB}{k^2 \hbar^2}\cot\theta_0 2k\alpha L_{arm} 
\]
The polarisation rotation in one of the arms of a spin-echo spectrometer, $\phi_{se} = 4\pi m\mu_n/h^2 \lambda B L_{arm}$ results in a spin-echo length of $\delta=2 m\mu_n/h^2 \lambda^2 BL_{arm} \cot \theta_0$  Hence, 
\[
\phi_e =  \delta 2k\alpha  .
\]
Unfortunately this value is only larger than the direct effect if the spin-echo length of the spectrometer is larger than $L$. As $L$ can easily be of the order of centimeters and the maximum spin-echo length is tens of micrometers, this is not a realistic enhancement opportunity. However, if the sample coherent scattering dimensions are much smaller than the spin-echo length, this is comparable to the normal SESANS scattering description. Hence, if one uses an electro-optical material in which the electric fields can be orders of magnitude larger than in vacuum $10^4$ to $10^5$~MV/m, it could be measured by means of this technique. If it would be an opportunity it would be essential that the splitting of the beam due to the triangular shaped electric field is in the same direction as the splitting of the beam in the spin-echo instrument. In the SESANS instrument the splitting occurs in the same direction as the guide field, hence the triangular shaped condensator plates can be positioned parallel to the guide field. For the OffSpec instrument the splitting occurs perpendicular to the guide field. Hence, before the triangular shaped condensator plates, the plane of rotation of the polarisation vector should be changed toward the effective magnetic field direction by an appropriate rotator. This is shown schematically in figure~\ref{fig:3}. 
After the field region the plane of rotation must be reversed to the initial spin-echo direction. 

Now, the question: 'In what way does this spin-echo length change affect the polarisation change of the beam?' can be answered. One could argue that this spin-echo length change can not be discriminated from a spin-echo length change due to a change in the instrument parameters of a spin-echo interferometer. This is supported by the optional construction of a spin-echo interferometer with triangular shaped magnetic field regions as described by Rekveldt~\cite{SESANS} and Keller~\cite{Keller1997}. Then, it can be corrected by changing the spin-echo length of one of the arms of the spin-echo interferometer or by an additional magnetic field produced by a phase coil. Another possibility to obtain the spin-echo polarisation rotation is to argue that the beam splitting on an inclined interface of a magnetic field region gives a refraction angle given by equation~(\ref{eq:refang}). This must compensate the refraction angle due to the electric field given by equation~(\ref{eq:eff}). The above derivation however suggests that neither of these assumptions is correct. A splitted beam in-itself gives only a very small amount of polarisation rotation. The SESANS effect can only be attributed to the fact that the scattering angle of the complete wave function changes, and hence traverses a changed length through the field regions in the second spin-echo arm and experiences a different rotation. The amount of rotation due to splitting or the angle between the propagation of the spin-up and spin-down wave function can be neglected and hence these effects can be ignored.
\section*{Conclusions}
With a neutron spin-echo spectrometer, it is possible to measure the polarisation phase change due to the influence of the electric field on the moving magnetic moment of the neutron. The use of a material with a high breakdown electric field could enhance the effect by two orders of magnitude. 
\\
\section*{Acknowledgments}
The author wishes to thank Theo M. Rekveldt for fruitful discussions.

\bibliography{main}

\end{document}